\documentclass{article}

\usepackage[utf8]{inputenc}
\usepackage{repsize}
\usepackage{latexsym}
\usepackage{defs}
\usepackage{amsmath,amssymb,amsfonts}
\usepackage{color}
\usepackage{graphics}

\newcommand{\remove}[1]{}

\newcommand{\PP}{\mmathp{\bf P}}
\newcommand{\maxsat}{{\sc max2sat{\tt(}3{\tt)}}}
\newcommand{\mvpp}{{\sc mvpp}}
\newcommand{\Vopt}{\ensuremath{V_{\rm opt}}}
\newcommand{\Mopt}{\ensuremath{M_{\rm opt}}}

\title{APX-Hardness of the Minimum Vision Points Problem}

\date{}

\author{
Mayank Chaturvedi\thanks{Birla Institute of Technology and Science Pilani, Goa Campus, India.\hfill\mbox{}email:~\texttt{\color{blue}f20170548@goa.bits-pilani.ac.in}
}
\and
Bengt J.~Nilsson\thanks{Department of Computer Science and Media Technology, Malmö University, Sweden.\hfill\mbox{} email:~{\tt\color{blue}bengt.nilsson.TS@mau.se}
}
}

\begin{document}

\maketitle

\begin{abstract}
Placing a minimum number of guards on a given watchman route in a polygonal domain is called the {\em minimum vision points problem}. We prove that finding the minimum number of vision points on a shortest watchman route in a simple polygon is APX-Hard. We then extend the proof to the class of rectilinear polygons having at most three dent orientations. 
\end{abstract}

\graphicspath{{Figures/}}

\section{Introduction}\label{intro} 
The problem of guarding polygonal domains is known as the {\em Art Gallery Problem}. A {\em guard\/} is a point in the domain and the visibility of the guard is defined to be those points that can be reached from the guard by line segments that do not intersect the exterior of the domain. 
When the domain is a simple polygon, Aggarwal~\cite{Aggarwal} and Lee and Lin~\cite{Lee} independently prove that finding the minimum number of guards is NP-hard; see also O'Rourke~\cite{ORourke}, this is later strengthened to APX-hardness and $\exists\mathbb{R}$-hardness~\cite{Abrahamsen,Broden-APX,Eidenbenz}.
    
Computing the watchman route is another way to solve the guarding problem. A {\em watchman route\/} is a closed tour traced by a moving guard who sees the complete polygon while tracing the tour. 
There exist polynomial time algorithms that compute the shortest watchman route for simple polygons~\cite{Dror-fixed,Tan-floating,Tan-fixed}.
    
Surveillance devices that trace the watchman route to guard a polygonal domain may be unable to accurately engage their vision systems continuously and could potentially only do so at discrete points along the tour. Such points are called {\em vision points}. The optimization problem of finding a minimum number of vision points on a shortest watchman route is denoted the {\em minimum vision points problem}~(\mvpp) and is the focus of our current results. The \mvpp\ has the potential of being computationally easier than the original art gallery problem and indeed \mvpp\ admits a polynomial time exact solution for straight-walkable polygons and street polygons~\cite{Carlsson-Nilsson} whereas the art gallery problem has been shown NP-hard already for monotone polygons~\cite{Krohn-Nilsson}, monotone polygons being a subset of straight-walkable polygons; see also Ghosh and Burdick~\cite{Ghosh} for results for polygons with holes. 

On the negative side, Carlsson~{\em et~al}.~\cite{Carlsson-NPHard} prove the NP-hardness of finding a minimum number of vision points on a shortest watchman route in a simple polygon and the claim by Abrahamsen~{\em et~al}.~\cite{Abrahamsen} that the art gallery problem is $\exists\mathbb{R}$-hard also for guards restricted to the boundary of a polygon further strengthens the complexity of \mvpp\ since it shows that the \mvpp\ is also $\exists\mathbb{R}$-hard in simple polygons. Given a simple polygon \PP\!, we can add small notches (consisting of a constant number of edges) at each of the convex vertices of \PP\!, thus creating a new polygon \PP'\ such that the shortest watchman route follows each edge of the original polygon~\PP\!, whereby the minimum set of vision points for \PP'\ is a superset of the minimum guard set of \PP\ restricted to the boundary.
    
The hardness results exhibited for these problems increases the importance of obtaining good approximation methods. 
We show that there is a limit to how well such approximation methods can be by showing that \mvpp\ is APX-hard for simple polygons and we also extend the result to also hold for rectilinear polygons having three dent orientations~\cite{Culberson,Motwani,Motwani-perfect}, thus likely excluding the existence of polynomial time approximation schemes for~\mvpp.
\section{A Reduction for Simple Polygons}\label{simple} 
We make a gap preserving reduction~\cite{Ausiello} from \maxsat\ to \mvpp\ in simple polygons, where \maxsat\ is the following problem.
\begin{quote}
    \maxsat\\
    {\sc Instance}: a set of $n$ boolean variables $u_1,\ldots,u_n$ and a set of $m$ clauses $c_1,\ldots,c_m$, each consisting of a disjunction of exactly two distinct literals formed from the $n$ variables such that each variable occurs at most three times in the clauses.\\
    {\sc Solution:} an assignment to the variables that satisfies the largest number of clauses.
\end{quote}

Berman and Karpinski~\cite{Berman} show that it is NP-hard to approximate \maxsat\ by a factor $2012/2011-\epsilon$, for any $\epsilon>0$.

Given an instance of \maxsat, we can assume that no variable occurs only non-negated or negated in the clauses, otherwise we simply assign it the appropriate truth value to satisfy those clauses it is contained in. Since a variable then occurs two or three times in the clauses, there is one version, non-negated or negated, that occurs exactly once. We call this the {\em lone literal\/} of the variable. We also assume that at least one variable occurs three times in the clauses and without loss of generality that $u_1$ is such a variable with $\bar{u}_1$ as the lone literal in clause $c_1$. This will be used later to argue the structure of a canonical set of vision points.

\begin{lemma}\label{lem:maxsat}
For an instance of \maxsat\ having $n$ variables and $m$ clauses, such that $M$ of them are satisfiable, it holds that: 
\begin{enumerate}
    \item $n\leq m$,
    \item $M\geq3m/4$,
	\item there exists an optimal solution such that any unsatisfied clause consists only of lone literals.
\end{enumerate}
\end{lemma}
\begin{proof}
Claim~1.\@~holds since each variable appears as at least two literals in the clauses.

For Claim~2., a random assignment will satisfy each clause with probability at least $3/4$ so in total $3m/4$ clauses are satisfied in expectation. Therefore, at least one assignment must exist having at least  $3m/4$ clauses satisfied.

For Claim~3., any unsatisfied clause that contains a non-lone literal can be satisfied by reversing the assignment of that variable. Only the clause  containing the lone literal can become unsatisfied by this operation so the number of satisfied clauses does not decrease.
\end{proof}
We call a solution to a \maxsat\ instance that obeys Claims~1.,~2.,~and~3.~in Lemma~\ref{lem:maxsat}, {\em locally maximal}.

\begin{figure}
\begin{center}
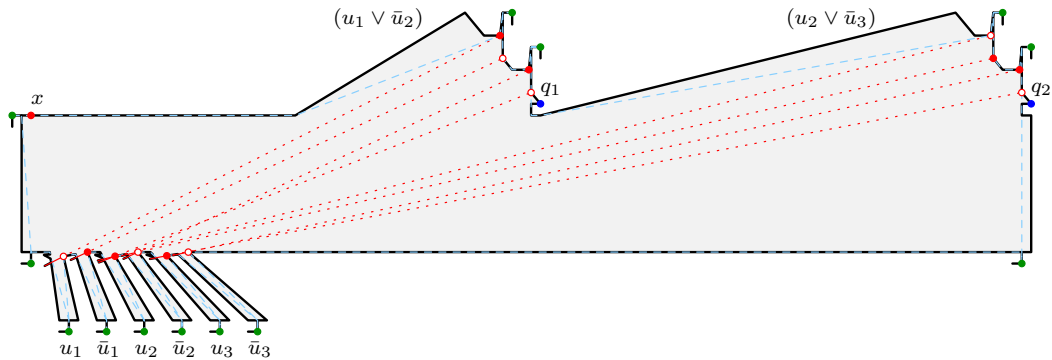
\end{center}
\caption{\label{vppredfig}The corresponding \mvpp\ instance for the \maxsat\ instance \((u_1\vee \bar{u}_2), (u_2 \vee \bar{u}_3)\).}
\end{figure}

\begin{figure}
\begin{center}
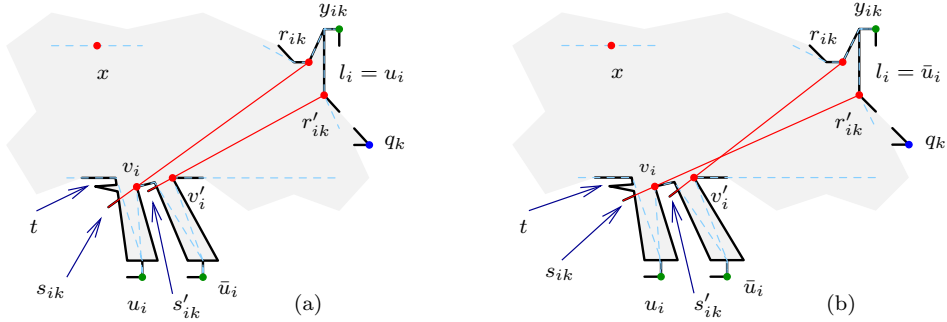
\end{center}
\caption{\label{setupfig}Connecting variables and clauses in the construction.}
\end{figure}

\medskip
To prove APX-Hardness of \mvpp\ in simple polygons, we reduce \maxsat\ to \mvpp. 
We modify the NP-hardness proof by Aggarwal~\cite{Aggarwal} and Lee and Lin~\cite{Lee} in the same way as Carlsson~{\em et al}.~\cite{Carlsson-NPHard}; see Figure~\ref{vppredfig}.%
The original NP-hardness proof constructs a {\em reduction polygon\/} for a \maxsat\ instance consisting of a base rectangle with clause gadgets along the upper segment and variable gadgets along the lower segment. Each clause gadget $c_k$ is a structure with two {\em chimneys\/} corresponding to the literals in the clause having a designated point $q_k$ that is seen using two guards if and only if the clause is satisfied. Each variable gadget consists of two {\em wells}, visible from the point $x$, one corresponding to the literal $u_i$ and the other corresponding to the literal $\bar{u}_i$. Each variable gadget has a spike $t$, the only vertices seeing $t$ being its adjacent vertices along the polygon boundary and the two vertices $v_i$ and $v'_i$ marked red in Figure~\ref{setupfig}. Each literal chimney in a clause $c_k$ has two red vertices $r_{ik}$ and $r'_{ik}$ that see it and they are connected to the corresponding variable gadget of $u_i$. For the clause $c_k=(l_i\vee l_{j})$ the literal chimney of $l_i$ is connected to variable gadget $u_i$ by adding spikes $s_{ik}$ and $s'_{ik}$ depending on whether $l_i$ is $u_i$ or $\bar{u}_i$ as illustrated in Figures~\ref{setupfig}(a) and~(b). At least one guard in the clause gadget $c_k$ must be placed at the lower vertex $r'_{ik}$ or $r'_{jk}$ to see both the chimney and the point $q_k$, the rightmost point of the clause gadget. The chimney is made thin enough so that no point sees the top $y_{ik}$ of more than one chimney; see Figures~\ref{vppredfig} and~\ref{setupfig}.%

To adapt the construction for \mvpp, we extend it by adding two {\em caves\/} (thin corridors each with a $90^{\circ}$ bend) to each clause structure, one at the top of each chimney, one cave is added at the bottom of each well structure, and three extra caves, on the top right sides of the well structures and one at~$x$, are also added. This gives us $2n+2m+3$ caves. To guard the polygon using vision points, the shortest watchman route must enter each of the caves and thus has a vision point in each cave, these are marked green in Figures~\ref{vppredfig} and~\ref{setupfig}. The caves tie down the shortest watchman route to ensure that the route passes the {\em critical guard points\/} that are used in the constructions of Aggarwal~\cite{Aggarwal} and Lee and Lin~\cite{Lee}. These critical guard points are marked red in Figures~\ref{vppredfig} and~\ref{setupfig}. The polygon and shortest watchman route have polynomial sized descriptions in the size of the \maxsat\ instance.

Disregarding the vision points in the caves, each clause gadget requires at least two more vision points (and at most three) and each variable gadget requires at least one vision point given a vision point at $x$. Thus, if an assignment to the \maxsat\ instance satisfies $M$ clauses, we can guard the polygon using 
$
V = 
2n+2m+3+n+2M+3(m-M)+1=3n+5m-M+4
$ 
vision points by placing one in each cave, one at $x$, one on the critical guard point corresponding to the assigned truth value in the variable gadget, one on the matching critical guard point of each literal chimney and, if the clause $c_k$ is not satisfied, one extra vision point at the rightmost of the critical guard points $r'_{ik}$ in the clause structure to see~$q_k$. We call such a placement a {\em canonical vision point set\/} and prove that we can assume that any vision point set is canonical. (A canonical vision point set is given in Figure~\ref{vppredfig} consisting of the cave guards in green, the point $x$ and the subset of red points that have white centers.)

\medskip%
Given a set of vision points, we modify it to be canonical without increasing its size as follows. Clearly each green point must be a vision point otherwise not all caves are seen. We can also assume that point $x$ is a vision point, otherwise each variable gadget must have two further vision points and, since each clause gadget must also have two further vision points, we obtain at least $4n+4m+3$ vision points. Without loss of generality, these are the critical guard points $v_i$ and $v'_i$ in the variable gadgets and $r'_{ik}$ and $r'_{jk}$ in each clause gadget $c_k=(l_i\vee l_{j})$, giving exactly $4n+4m+3$ vision points guarding the polygon. Since $u_1$ occurs in three clauses and has $\bar{u}_1$ as lone literal in clause $c_1$, we remove the vision point at $v'_1$, place it at $x$, and move the vision point at $r'_{1,1}$ to $r_{1,1}$ if necessary, thus neither decreasing coverage nor increasing the size of the vision point set.

The top point $y_{ik}$ of a clause gadget chimney of $l_i$ in clause $c_k$ sees two connected components of the watchman route that we denote $w$ and $w'$, $w$ containing $y_{ik}$. The component $w$ contains the chimney's two critical guard points $r_{ik}$ and $r'_{ik}$, $r'_{ik}$ seeing $q_{k}$. If $w'$ contains vision points, we move them to $r'_{ik}$, if the path from $y_{ik}$ to $r_{ik}$ of $w$ contains vision points, we move them to $r_{ik}$, and if the path from $y_{ik}$ to $r'_{ik}$ of $w$ contains vision points, we move them to $r'_{ik}$. If $r'_{ik}$ has a vision point after these moves, we remove all other vision points that see $y_{ik}$ (except the green cave one), otherwise we keep one at $r_{ik}$. Together with $x$, this vision point will guard at least as much as the original vision points on $w$ and~$w'$ (except for a disregardable portion of the other literal chimney in the clause gadget).

The apex of the spike $t$ in a variable gadget $u_i$ sees three connected components of the watchman route. We denote these by $w_1$, $w_2$, and $w_3$ in increasing order of distance to $t$ and note that $w_2$ contains critical guard point $v_i$ and $w_3$ contains $v'_i$. If $w_1$ or $w_2$ have vision points, we move them to $v_i$ and if $w_3$ has vision points, we move them to $v'_i$ and remove any duplicates from $v_i$ and $v'_i$. Together with $x$, these points will guard at least as much as the original vision points on $w_1$, $w_2$, and $w_3$. If both $v_i$ and $v'_i$ have vision points, we remove the one that corresponds to the lone literal in the clause gadget of some clause $c_k$ and place one at $r'_{ik}$ unless point $q_k$ is already seen by the other vision points in the clause gadget. 
The process described above never adds vision points so the size of a canonical vision point set 
is no larger than the original set.
We state this as a lemma.
\begin{lemma}\label{lem:canonical}
Any vision point set on a shortest watchman route in a reduction polygon can be transformed to a canonical vision point set of no larger size than the original set.
\end{lemma}

Berman and Karpinski~\cite{Berman} show that it is NP-hard to approximate \maxsat\ by a factor $2012/2011-\epsilon$, for any $\epsilon>0$. 
Assume from the discussion above that we have a polynomial time approximation algorithm for \mvpp\ that produces $V=3n+5m-M+4$ canonical vision points for some value $M$.
We can assume that $M$ corresponds to some locally maximal solution of the \maxsat\ instance for which the optimum is \Mopt. 
Given an optimal solution to the \maxsat\ instance, 
we construct a canonical vision point set in the reduction polygon by assigning vision points according to the truth values in the \maxsat\ solution. Let $V'=3n+5m-\Mopt+4$ be the number of vision points placed in this way in the reduction polygon and let \Vopt\ be the minimum number of vision points in the reduction polygon. Since $V'\geq\Vopt$, $\Mopt/M\geq2012/2011-\epsilon$, and $m\geq4$,
we
have by Lemmata~\ref{lem:maxsat} and~\ref{lem:canonical} the ratio
\begin{align}
\frac{V}{\Vopt}
& \geq
\frac{V}{V'}
=
\frac{
3n+5m-M+4
}{
3n+5m-\Mopt+4
}
\geq
\frac{
9m-M
}{
9m-M(2012/2011-\epsilon)
}
\geq
\frac{
22121
}{
22120
}-\delta,
\end{align}
for any $\delta>0$ dependent on~$\epsilon$, which proves the APX-hardness of \mvpp\ in simple polygons.

We have proved the following theorem.
\begin{theorem}
For every $\delta>0$, it is NP-hard to approximate \mvpp\ in a simple polygon to within a factor of\/~$22121/22120-\delta$.
\end{theorem}


\section{The \mvpp\ in Rectilinear Polygons with Three Dents}\label{threedent} 

\begin{figure}
\begin{center}
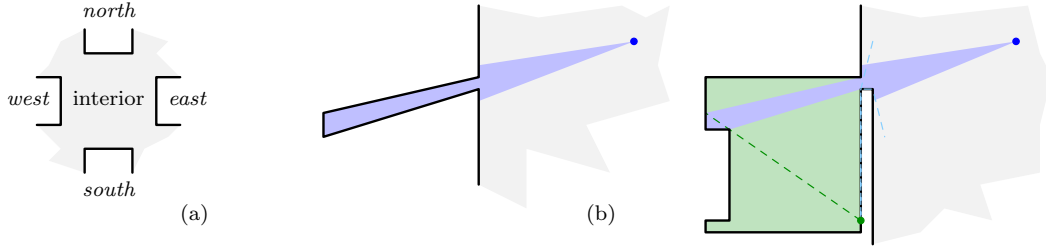
\end{center}
\caption{\label{dentfig}Illustrating the concepts of dents and the rectilinear spike emulator.}
\end{figure}

The concept of {\em dents\/} in rectilinear polygons was introduced by Culberson and Reckhow~\cite{Culberson} and Motwani~{\em et~al}.~\cite{Motwani,Motwani-perfect} and they develop algorithms for orthogonal covering problems in rectilinear polygons with restricted number of dent orientations. A {\em dent\/} in a rectilinear polygon is simply a boundary edge where both endpoints are reflex. Thus, we identify dents with four different orientations, {\em north}, {\em south}, {\em east}, and {\em west\/}; see Figure~\ref{dentfig}(a).

Monotone rectilinear polygons have dents of one or two (opposite) orientations and for these, optimal linear time algorithms for \mvpp\ exist~\cite{Carlsson-Nilsson,Carlsson-NPHard}. We settle the complexity status for polygons with three dent orientations here but for rectilinear polygons having two (non-opposite) dent orientations the complexity status remains unknown. This should be contrasted with the classical art gallery problem, where linear time algorithms for computing the minimum number of point guards exist only for rectilinear polygons having one dent orientation (histograms)~\cite{Carlsson-NPHard}, for rectilinear polygons having two non-opposite dent orientations, the art gallery problem can be shown to be APX-hard by modifying the proof by Brodén~{\em et~al}.~\cite{Broden-APX} slightly. For the classical art gallery problem, the complexity is unknown for rectilinear monotone polygons having two opposite dent orientations but we suspect that the problem is indeed NP-hard given the recent NP-hardness proof for vertex guarding rectilinear staircase polygons~\cite{Gibson-staircase}.

\medskip
We modify the reduction introduced in the previous section to be rectilinear and furthermore to only contain dents of three different orientations. To this end, we introduce the {\em rectilinear spike emulator}, also used by Katz and Roisman~\cite{Katz}. A spike as used in Section~\ref{simple} is a thin corridor that can only be seen along a thin visibility cone. We can emulate the effect with a rectilinear gadget as shown in Figure~\ref{dentfig}(b) using one extra guard (green in the figure) and, as long as the original spike has reflex vertices with larger $x$-coordinates than its convex vertices, the rectilinear spike gadget never introduces an east dent.
\begin{figure}
\begin{center}
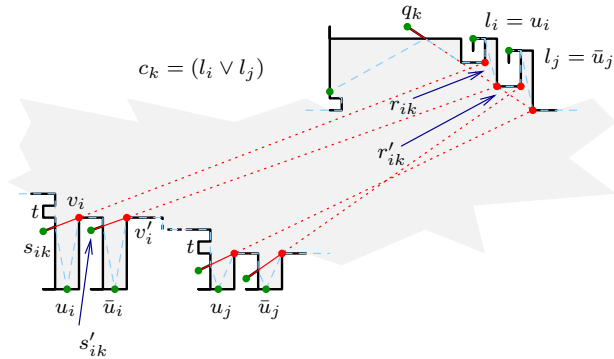
\end{center}
\caption{\label{rectilineargadgetfig}Illustrating the variable and clause gadgets in the rectilinear construction. Spikes at $s_{ik}$, $s'_{ik}$, and $q_{k}$ are replaced by small rectilinear spike emulators.}
\end{figure}

As in Section~\ref{simple}, each variable gadget consists of two rectilinear wells, 
corresponding to the literals  $u_i$ and $\bar{u}_i$. The point $x$ is not necessary, since each well is covered by a green guard at the bottom.
Each variable gadget has a rectilinear spike $t$ seen by the two critical guard points $v_i$ and $v'_i$ marked red in Figure~\ref{rectilineargadgetfig}. As before, each clause gadget has two rectilinear chimneys corresponding to the literals in the clause and each chimney in a clause $c_k$ has two critical guard points $r_{ik}$ and $r'_{ik}$ that see it and they are connected to the corresponding variable gadget of $u_i$ by adding rectilinear spike emulators $s_{ik}$ and $s'_{ik}$ as illustrated in Figure~\ref{rectilineargadgetfig}. Again, we note that at least one guard in a clause gadget $c_k=(l_i\vee l_{j})$ must be placed at the lower vertex $r'_{ik}$ or $r'_{jk}$ to see both the chimney and the point $q_k$, placed in rectilinear spike emulator at the top edge of the clause gadget; see Figure~\ref{rectilineargadgetfig}.%

\begin{figure}
\begin{center}
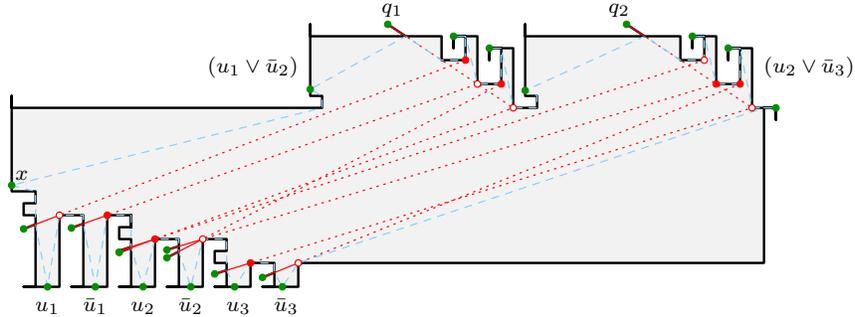
\end{center}
\caption{\label{rvppredfig}A rectilinear \mvpp\ instance for the \maxsat\ instance \((u_1\vee \bar{u}_2), (u_2 \vee \bar{u}_3)\).}
\end{figure}

We add caves at the top of the chimneys, at the bottom of the variable gadgets, on the right side of the base rectangle and two caves, ensuring that these do not introduce east dents. These tie down the shortest watchman route to make it pass all the critical guard points. The convex vertices of the shortest watchman route each require a vision point, giving us $2n+8m+2$ such green vision points. In the same way as in Section~\ref{simple}, we can argue that any algorithm produces a canonical vision point set consisting of
$
V = 
2n+8m+2+n+2M+3(m-M)
=
3n+11m-M+2
$ 
vision points, choosing the remaining ones from the set of critical guard points; see Figure~\ref{rvppredfig} for a full example of a canonical vision point set consisting of the green points and the subset of the red points that have white centers in a rectilinear polygon with three dent orientations.

\medskip%
Using the result by Berman and Karpinski~\cite{Berman}, that 
it is NP-hard to approximate \maxsat\ by a factor $2012/2011-\epsilon$, for any $\epsilon>0$, 
we obtain as before the ratio
\begin{align}
\frac{V}{\Vopt}
& \geq
\frac{
3n+11m-M+2
}{
3n+11m-\Mopt+2
}
\geq
\frac{
15m-M
}{
15m-M(2012/2011-\epsilon)
}
\geq
\frac{
38209
}{
38208
}-\delta,
\end{align}
for any $\delta>0$ dependent on $\epsilon$,
proving the APX-hardness of \mvpp\ in rectilinear polygons having three dent orientations.

We have proved the following theorem.
\begin{theorem}
For every $\delta>0$, it is NP-hard to approximate \mvpp\ in a simple rectilinear polygon with three dent orientations to within a factor of\/~$38209/38208-\delta$.
\end{theorem}

\bibliography{Bibliography}

\begin{thebibliography}{10}

\bibitem{Abrahamsen}
M.~Abrahamsen, A.~Adamaszek, and T.~Miltzow.
\newblock The art gallery problem is $\exists\mathbb{R}$-complete.
\newblock {\em Journal of the ACM}, 69(1):1--70, 2021.

\bibitem{Aggarwal}
A.~Aggarwal.
\newblock {\em The Art Gallery Theorem: It's Variations, Applications and
  Algorithmic Aspects}.
\newblock PhD thesis, Department of Electrical Engineering and Computer
  Science, Johns Hopkins University, 1984.

\bibitem{Ausiello}
G.~Ausiello, A.~Marchetti-Spaccamela, P.~Crescenzi, G.~Gambosi, M.~Protasi, and
  V.~Kann.
\newblock {\em Complexity and Approximation --- Combinatorial Optimization
  Problems and Their Approximability Properties}.
\newblock Springer, 1999.

\bibitem{Berman}
P.~Berman and M.~Karpinski.
\newblock On some tighter inapproximability results.
\newblock In {\em Proc.~$26^{\rm th}$ International Colloquium on Automata,
  Languages and Programming, ICALP'99}, pages 200--209, 1999.

\bibitem{Broden-APX}
B.~Brodén, M.~Hammar, and B.J. Nilsson.
\newblock Guarding lines and 2-link polygons is {APX}-hard.
\newblock In {\em Proc.~$13^{\rm th}$ Canadian Conference on Computational
  Geometry, CCCG'01}, pages 45--48, 2001.

\bibitem{Carlsson-Nilsson}
S.~Carlsson and B.J.~Nilsson.
\newblock Computing vision points in polygons.
\newblock {\em Algorithmica}, 24(1):50--75, 1999.

\bibitem{Carlsson-NPHard}
S.~Carlsson, B.J.~Nilsson, and S.~Ntafos.
\newblock Optimum guard covers and $m$-watchmen routes for restricted polygons.
\newblock {\em International Journal of Computational Geometry and
  Applications}, 3(1):85--105, 1993.

\bibitem{Culberson}
J.C.~Culberson and R.A.~Reckhow.
\newblock Orthogonally convex coverings of orthogonal polygons without holes.
\newblock {\em Journal of Computer and System Sciences}, 39:166--204, 1989.

\bibitem{Dror-fixed}
M.~Dror, A.~Efrat, A.~Lubiw, and J.S.B.~Mitchell.
\newblock Touring a sequence of polygons.
\newblock In {\em Proc.~35th ACM Symposium on Theory of Computing, STOC'03},
  pages 473--482, 2003.

\bibitem{Eidenbenz}
S.~Eidenbenz.
\newblock Inapproximability results for guarding polygons without holes.
\newblock In {\em Proc.~$9^{\rm th}$ International Symposium on Algorithms and
  Computation,~ISAAC'98}, volume LNCS~1533, pages 427--437. Springer, 1998.

\bibitem{Ghosh}
S.K.~Ghosh and J.W.~Burdick.
\newblock An on-line algorithm for exploring an unknown polygonal environment
  by a point robot.
\newblock In {\em Proc.~$9^{\rm th}$ Canadian Conference on Computational
  Geometry, CCCG'97}, pages 100--106, 1997.

\bibitem{Gibson-staircase}
M.~Gibson-Lopez, E.~Krohn, B.J.~Nilsson, M.~Rayford, S.~Soderman, and P.~\.Zyli\'nski
\newblock On Vertex Guarding Staircase Polygons.
\newblock Submitted to the 15th Latin American Theoretical Informatics Symposium LATIN'22,~2022.

\bibitem{Katz}
M.J.~Katz and G.S.~Roisman.
\newblock On guarding the vertices of rectilinear domains.
\newblock {\em Computational Geometry}, 39(3):219--228, 2008.

\bibitem{Krohn-Nilsson}
E.A.~Krohn and B.J.~Nilsson.
\newblock The complexity of guarding monotone polygons.
\newblock In {\em Proc.~24th Canadian Conference on Computational Geometry,
  CCCG'2012}, pages 167--172, 2012.

\bibitem{Lee}
D.T.~Lee and A.K.~Lin.
\newblock Computational complexity of art gallery problems.
\newblock {\em IEEE Transactions on Information Theory}, IT-32:276--282, 1986.

\bibitem{Motwani}
R.~Motwani, A.~Raghunathan, and H.~Saran.
\newblock Covering orthogonal polygons with star polygons: The perfect graph
  approach.
\newblock {\em Journal of Computer and System Sciences}, 40:19--48, 1989.

\bibitem{Motwani-perfect}
R.~Motwani, A.~Raghunathan, and H.~Saran.
\newblock Perfect graphs and orthogonally convex covers.
\newblock {\em {SIAM} Journal on Algebraic Discrete Methods}, 2:371--392, 1989.

\bibitem{ORourke}
J.~O'Rourke.
\newblock {\em Art Gallery Theorems and Algorithms}.
\newblock Oxford University Press, 1987.

\bibitem{Tan-floating}
X.-H.~Tan.
\newblock Fast computation of shortest watchman routes in simple polygons.
\newblock {\em Information Processing Letters}, 77(1):27--33, 2001.

\bibitem{Tan-fixed}
X.-H.~Tan, T.~Hirata, and Y.~Inagaki.
\newblock Corrigendum to ``an incremental algorithm for constructing shortest
  watchman routes''.
\newblock {\em International Journal of Computational Geometry and
  Applications}, 9(3):319--324, 1999.

\end{thebibliography}

\end{document}